\documentclass[epjST]{svjour}
\usepackage{graphicx}
\usepackage{amsmath,amsfonts,amssymb}
\begin{document}
\title{L\'evy Ratchet in a Weak Noise Limit:  Theory and Simulation}
\author{I.\ Pavlyukevich\inst{1}         \fnmsep\thanks{\email{ilya.pavlyukevich@uni-jena.de}}
\and B.\ Dybiec \inst{2}                      \fnmsep\thanks{\email{bartek@th.if.uj.edu.pl      }}
\and A.\ V.\ Chechkin \inst{3}              \fnmsep\thanks{\email{achechkin@kipt.kharkov.ua  }}
\and I.\ M.\ Sokolov\inst{4}                 \fnmsep\thanks{\email{igor.sokolov@physik.hu-berlin.de }}
}
\institute{Institut f\"ur Stochastik, Friedrich--Schiller--Universit\"at Jena, Ernst--Abbe--Platz 2, 07743 Jena, Germany
\and M.\ Smoluchowski Institute of Physics, and
M.\ Kac Center for Complex Systems Research, Jagellonian University, ul.\ Reymonta 4, 30--059 Krak\'ow, Poland
\and School of Chemistry, Tel Aviv University, Ramat Aviv, Tel Aviv 69978, Israel and Institute for Theoretical Physics NSC KIPT,
Akademicheskaya st.\ 1, Kharkov 61108, Ukraine
\and Institut f\"ur Physik, Humboldt--Universit\"at zu Berlin, Newtonstrasse 15, 12489 Berlin, Germany
}
\abstract{
We study the motion of a particle embedded in a time independent
periodic potential with broken mirror symmetry and subjected to a
L\'evy noise possessing L\'evy stable probability law (L\'evy
ratchet). We develop analytical approach to the problem based on the
asymptotic probabilistic method of decomposition proposed by P.\
Imkeller and I.\ Pavlyukevich [J.\ Phys.\ A {\bf39}, L237 (2006);
Stoch.\ Proc.\ Appl.\ {\bf116}, 611 (2006)]. We derive analytical
expressions for the quantities characterizing the particle motion,
namely the splitting probabilities of first escape from a single
well, the transition probabilities and the particle current. A
particular attention is devoted to the interplay between the
asymmetry of the ratchet potential and the asymmetry (skewness) of
the L\'evy noise. Intensive numerical simulations demonstrate a good
agreement with the analytical predictions for sufficiently small
intensities of the L\'evy noise driving the particle.
} 
\maketitle

\section{Introduction\label{sec:introduction}}

L\'evy motion, also referred to as ``L\'evy flights'', stands for a
class of non-Gaussian Markovian random processes whose stationary
increments are distributed according to the L\'evy stable
probability laws originally studied by French mathematician Paul
Pierre L\'evy \cite{levy}. The term ``L\'evy flights'' (LF) was coined by
Mandelbrot \cite{mandelbrot}, who thus poeticized this type of
random motion which is now considered as a paradigm of non-Brownian
random walk. Similar to Brownian motion, LFs have a solid
probabilistic background. Indeed, the central limit theorem and the
properties of Gaussian probability laws and processes constitute
mathematical foundation of the Brownian motion \cite{feller}. At the
same time, the \emph{generalized} central limit theorem and the
remarkable properties of the L\'evy stable probability laws serve as
the basis of the LFs theory. Generalized central limit theorem says
that the stable probability laws, like the Gaussian law, attract the
distributions of sums of random variables
\cite{gnedenko-kolmogorov}. Due to this reason, L\'evy stable distributions naturally
appear when evolution of a system, or result of an experiment are
determined by the sum of independent, identically distributed random
factors. The probability density functions (PDFs) of the stable laws
exhibit slowly decaying, power-law asymptotic behavior of the form
$|x|^{-(1+\alpha)}$, where $\alpha$ is called the L\'evy index, $0 <
\alpha < 2$.  Due to this reason the stable PDFs appear naturally in
the description of random processes with large outliers, far from
equilibrium. Another important property of LFs is their statistical
self-similarity, or self-affinity \cite{samorod}. Therefore, like
the Brownian motion, they are naturally suited for the description
of random fractal processes. LFs are ubiquitous in nature and they
were observed in various fields of science, including physics
(stochastic dynamics, turbulent flows
\cite{nature93,phystoday96,swinney,diego1}, turbulence and
turbulent transport in magnetized plasmas \cite{chechkinplasma,goncharplasma,goncharplasma1,diego2}, optical lattices \cite{walther}, molecular
collisions \cite{galgani}), biology (heartbeat dynamics \cite{peng},
firing of neural networks \cite{segev}, searching on a folding
polymer \cite{lomholt}, foraging movement \cite{buldyrev}),
seismology (recording of seismic activity \cite{vidal}), stochastic
climate dynamics \cite{ditlevsen}, engineering (signal processing
\cite{kassam,WegmanTS88,NikiasS95}), economics (financial time series \cite{MantegnaS99,BouchaudP00,GomoryM97}),
and even the spreading of diseases and dispersal of banknotes
\cite{brockmann}. The list above is far from being complete; we only
mention that recently illuminating experiments with LFs of light
have been reported \cite{barthelemy,mercadier}.

The behavior of LFs in external fields brings a few surprising
effects. The PDF in a harmonic potential evolves to a stationary state
given by a L\'evy stable distribution \cite{Jespersen,CheGo} with
the same stability index  $\alpha$ as the underlying noise. For
unimodal power law potentials steeper than a harmonic one
$U(x) \propto |x|^c$, $c > 2$, the stationary state is
characterized by the asymptotic power law decay, $p(x) \propto |x|^{-(c+\alpha-1)} $.
These stationary probability densities decay  faster \cite{chemphys,SamorodnitskyG-03,SamGri07,dubkov}
than the corresponding L\'evy stable
density. Moreover, the stationary PDFs in such potentials
are bimodal, i.e.\ they possess a local minimum at the origin
and two maxima at $x\ne 0$. The study of unimodal to multimodal
bifurcations in this system during relaxation was addressed in
Ref.~\cite{chekla,ChechkinGKM-04,Ditlevsen-04}.
For subharmonic potential  $c < 2$ the decay $p(x) \propto |x|^{-(c+\alpha-1)} $ is slower than the decay of the
corresponding L\'evy stable density. Moreover, stationary states only exist for $c > 2-\alpha$ \cite{JStat}.

The properties of the LFs in external fields were studied with the
use of space-fractional Fokker-Planck equation and the Langevin
equation for a particle driven by L\'evy noise.
Motivated by the model from stochastic climate dynamics \cite{ditlevsen}, the barrier
crossing problem for LFs in generic types of the
potentials was studied in Refs.~\cite{ditlevsen_2,chegoslyu,ChechkinSMK-07,bartek_1,impavl_a,impavl_b}. In particular, the method of
decomposition of the L\'evy process into bounded jump component and compound
Poisson part proposed in Refs.~\cite{impavl_a,impavl_b} allowed for treating the escape problem analytically
(at least in the weak noise asymptotics), while
solving the corresponding space-fractional Fokker-Planck
equation poses considerable difficulties \cite{chegoslyu}. Further, the idea of decomposition was
applied to describe a metastable behavior of a multi-well dynamical
system driven by L\'evy noise \cite {ESAIM}. This have led to a new approach to fast simulated annealing problem
\cite{hartley} of efficient non-local search of
the deepest potential well \cite{pavl_jpa2007,pavl_jcp2007}. We will discuss the
decomposition method in more detail below, see
Sec.~\ref{sec:decomposition}.

L\'evy processes lead to a richer behavior than Gaussian processes to which they
reduce for $\alpha=2$. Moreover the limiting distributions for sums of independent identically
distributed random variables do not need to be symmetric.
The asymmetry of the noise can induce preferred direction
of the motion \cite{diegoche,dybiec2008e} and asymmetry of stationary states
\cite{dybiec2007d}. This in turn is responsible for occurrence of the dynamical
hysteresis \cite{dybiec2007e}.

In this paper we study, both analytically and numerically, LFs in a
ratchet potential, i.e.\ in a periodic potential lacking reflection
symmetry. Such spatial symmetry breaking gives rise to a net
directed transport in presence of non-equilibrium fluctuations. The
motion in ratchet potentials attracted considerable attention due to
its role in fluctuation-driven transport, see for example
Ref.~\cite{reimann} and references therein. By now, the physics of
``Brownian motors'' is well understood and the theory is well
developed. However, not too much is known about L\'evy ratchets.
In Ref.~\cite{bartek_2} the authors solved numerically the Langevin
equation with a white L\'evy noise for an overdamped particle in a
ratchet potential. They demonstrated the appearance of directional
transport using such measures of directionality as the
position of the median of particle's displacements distribution
characterizing the group velocity, and the interquantile distance
giving the distributions' width. Such an approach
allows the authors to study L\'evy ratchet in the whole range of the
L\'evy index $\alpha$. In Ref.~\cite{diegoche} the analysis was
restricted to the region $1<\alpha<2$, at the same time the
numerical analysis of the Langevin equation is complemented by
numerical solution of space-fractional Fokker-Planck equation.

The main goal of the present paper is to develop analytical theory
of the L\'evy ratchet and to verify its predictions by numerical
simulations. We have found that the probabilistic method of
decomposition successfully applied to the escape problem for LFs in
Refs.~\cite{impavl_a,impavl_b} also works for the L\'evy ratchet
description. As potential applications we can mention, following
Ref.~\cite{diegoche}, a ratchet-like transport of impurities in
magnetically confined fusion plasmas \cite{benkadda}, and a ratchet
transport in biology, since biological systems are intrinsically far
from equilibrium \cite{phillips}.

The rest of the paper is organized as follows. In
Section~\ref{sec:model} we formulate the L\'evy ratchet model. In
Section~\ref{sec:analytics} we give the essentials of the
probabilistic decomposition method and present the results of
analytical theory. In Section~\ref{sec:results} we compare
analytical results with the results of numerical simulations. The
conclusions and summary are presented in Section~\ref{sec:summary}.

\section{Model of the L\'evy ratchet\label{sec:model}}

\begin{figure}[ht]
\begin{center}
{\includegraphics[width=0.7\columnwidth]{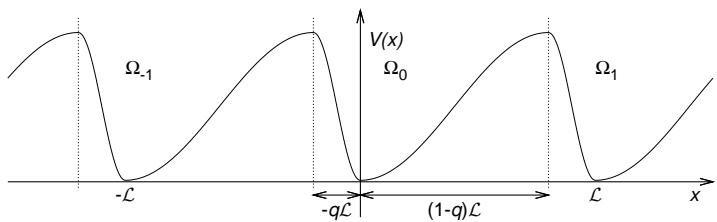}
}
\caption{Exemplary potential used for inspection of the ratchet
problem. The length of the potential segment is controlled by
$\mathcal{L}$ while the asymmetry of the potential is controlled by the
potential asymmetry parameter $q$ ($0<q<1$). For $q=1/2$ the
potential is symmetric.}
\label{fig:potential}
\end{center}
\end{figure}

We study the following Langevin equation for an overdamped particle
driven by the L\'evy stable noise $\dot L=dL/dt$
\begin{equation}
\dot x(t)=-V'(x)+\varepsilon \dot L(t),
\label{eq:langevin}
\end{equation}
where $V(x)$ is the potential, the prime denotes spatial derivative,
and $\varepsilon$ is the amplitude of the noise,
which will be considered to be small. The potential $V(x)$
is periodic, $V(x)=V(x+\mathcal{L})$, with potential wells
$\Omega_j$, $j=0,\pm1,\pm2,\dots$. The $j$-th potential well $\Omega_j$ is
located at
$x \in (-\mathcal{L}q+j\mathcal{L},\mathcal{L}(1-q)+j\mathcal{L})$
where $0<q<1$, see Fig.~\ref{fig:potential}. As an exemplary
potential $V(x)$ we use
\begin{equation}
V(x)=
\left\{
\begin{array}{ll}
V_0\left[ 1 - \cos\frac{\pi x}{a_1} \right], & 0 \leqslant x < a_1, \\
V_0\left[ 1 + \cos\frac{\pi(x - a_1)}{a_2} \right], & a_1 \leqslant x < \mathcal{L},
\label{eq:potential}
\end{array}
\right.
\end{equation}
where $\mathcal{L}=a_1+a_2$, $a_1=(1-q)\mathcal{L}$ and
$a_2=q\mathcal{L}$. The potential lacks reflection symmetry when
$q\neq1/2$. Compared with the $V_0\left[\sin(2\pi
x/\mathcal{L})+\sin(4\pi x/\mathcal{L})/4\right]$ potential typically used in
literature, e.g.\ \cite{reimann}, Eq.~(\ref{eq:potential}) offers an
advantage of an easier control of the spatial asymmetry. In all the
calculations presented below we use $V_0=\mathcal{L}=1$.
Figure~\ref{fig:potential} presents exemplary potential profile
given by Eq.~(\ref{eq:potential}).

In the integral form, Eq.~(\ref{eq:langevin}) reads
\begin{equation}
 x(t)=x(0)-\int_0^tV'(x(s))ds+\varepsilon L(t),
\label{eq:integrated}
\end{equation}
where $L(t)$ is the L\'evy stable motion, that is the (formal)
integral over the L\'evy stable noise in time.  For $\alpha\neq1$, the
Fourier-transform of $L(t)$, see Eqs.~(\ref{eq:langevin}) and~(\ref{eq:integrated}), is
\begin{equation}
\langle e^{ik L(t)}\rangle = \exp\left[
-tc|k|^\alpha\left(1-i\beta\mathrm{sgn}
k\tan\frac{\pi\alpha}{2}\right) \right],
\label{eq:charact}
\end{equation}
where $\alpha\in(0,2]$ is the stability index, $\beta\in[-1,1]$ is
the asymmetry (skewness) parameter, while $c$ is a positive
constant, $c^{1/\alpha}$ being called a scale parameter.
Initially, a
particle starts its motion in the minimum of the potential, e.g.\ in
the 0-th potential well. Furthermore, we assume $\beta\neq \pm 1$.

\section{Analytical approach to the L\'evy ratchet\label{sec:analytics}}

Analytical approach to the transport in a ratchet potential under
action of L\'evy noise is based on the decomposition method, for
which the main results were presented in ~\cite{impavl_a}, whereas
the mathematical details including proofs of corresponding theorems
were given in~\cite{impavl_b}. Then, the dynamics in multi-well
potentials was considered in~\cite{ESAIM}. Our exposition of the
L\'evy ratchet is based essentially on these three papers. In
Section~\ref{sec:analytics} we first give an ``intuitive''
explanation of the decomposition method and then present theoretical
results for main characteristics of ratchet transport, namely the
mean exit time from a single well, a transition probability, a mean
displacement and the current.

\subsection{Method of decomposition and single-well dynamics in case of symmetric L\'evy noise\label{sec:decomposition}}

Let us consider the discretized version of Eq.~(\ref{eq:langevin})
with the time step of integration $\Delta t$,
\begin{equation}
 x(n+1)-x(n)=-V'(x(n))\Delta t + \xi(n) ,
\label{eq:discrete Langevin}
\end{equation}
where $\xi(n)=L(n\Delta t)-L((n-1)\Delta t)=\varepsilon(c\Delta t)^{1/\alpha}\zeta(n)$ is the
value of the noise variable, or the L\'evy jump, at the $n$-th
interval of integration. Here, $\zeta(n)$ is the time-discrete L\'evy noise whose
characteristic function is given by Eq.~(\ref{eq:charact}) with $t =
1$, $c = 1$. Thus, the PDF of these L\'evy jumps $p=p(\xi)$ is
 characterized by the scale parameter, or characteristic length $w = \varepsilon (c \Delta
t)^{1/\alpha}$ depending on the parameters of the noise and on the
length $\Delta t$ of the time interval.

For the further exposition it is necessary to recall the asymptotics of the tails of the PDF $p(\xi)$ as $\xi\to\pm\infty$ given e.g.\ in
\cite[Chapter 4.3]{zolotarev_uchaikin}:
\begin{equation}
\label{eq:density}
p(\xi)\simeq c\Delta t \varepsilon^\alpha \frac{1\pm\beta}{2\cos(\frac{\alpha\pi}{2})|\Gamma(-\alpha)|}\frac{1}{\xi^{1+\alpha}}
=A_\pm w^\alpha\frac{1}{|\xi|^{1+\alpha}} ,\quad\xi\to\pm \infty, \ |\xi|\gg w
\end{equation}

\begin{figure}[ht]
\begin{center}{}
\includegraphics[width=0.5\columnwidth]{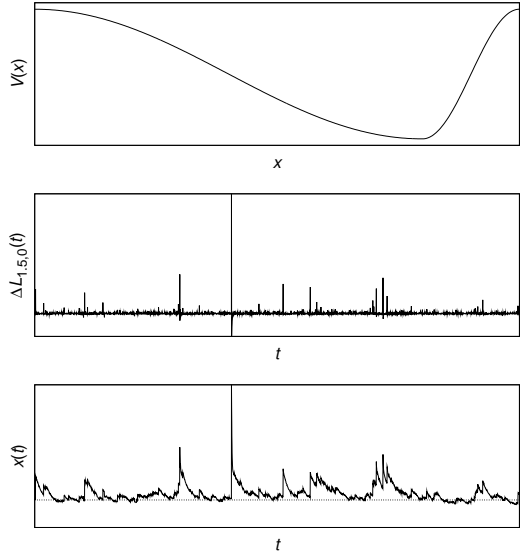}
\caption{Sample realization of the escape problem (bottom panel) driven by the L\'evy noise with parameters $\alpha=1.1$ and $\beta=0.9$
(middle panel). A particle moves in a potential well $\Omega_0$ (top panel).
}
\label{fig:decomposition}
\end{center}
\end{figure}

Let us  fix some level $\delta > 0$ (depending on $\varepsilon$) and
decompose the noise $\xi(n)$ into two parts: Thus all $\xi(n)$
smaller in absolute value than $\delta$ are considered to belong to
the background part of the noise (containing most of $\xi(n)$), and
large spikes with $|\xi(n)| > \delta$ form a shot-like noise
containing rare events, or outliers, see
Fig.~\ref{fig:decomposition} showing the decomposition
schematically. Since the values $\xi(n)$ are independent, the
probability that a spike occurs on the step $n$ does not depend
on $n$ and equals to
$P_{\text{spike}} = \int_{-\infty}^{-\delta} p(\xi)d\xi + \int_{\delta}^{\infty} p(\xi)d\xi $. For $\varepsilon$ and
$\Delta t$ small enough the characteristic length $w$ is small, thus
the probability $P_{\text{spike}}$ is determined by the large argument asymptotics
of the L\'evy stable density given by Eq.~\eqref{eq:density},
$p(\xi) \simeq A_\pm w^\alpha|\xi|^{-(1+\alpha)}$, $\xi\to\pm\infty$,
and behaves as $P_{\text{spike}} \simeq
(A/\alpha)c\Delta t \varepsilon^\alpha \delta^{-\alpha}$, where
$A=A_-+A_+$. Therefore, the spikes
form a Poissonian sequence of events, and, for $\delta$ large
enough, the spikes are well separated in time. The probability to
have at least one spike on the unit time interval is
\begin{equation}
P = \frac{A c}{\alpha}\varepsilon^{\alpha}\delta^{-\alpha}.
\label{probability}
\end{equation}
The probability not to have a spike on a time interval $[0,t]$ is around
$(1- (A c/\alpha)\varepsilon^\alpha \delta^{-\alpha}\Delta t
)^{t/\Delta t} \simeq \exp(-t/T)$ with
\begin{equation}
T = \frac{\alpha}{Ac}\varepsilon^{-\alpha}\delta^{\alpha}
\label{mean time}
\end{equation}
being the mean time between two subsequent spikes. On the other hand,
the background part of the noise has finite variance $\sigma^2$ and
its action over time intervals much larger than $\Delta t$ can be
modelled by the action of a white Gaussian noise. The variance
of the background component is given by
 $\sigma^2=\text{var}(\xi(n)\big|\,|\xi(n)|\leqslant \delta)$ and, for large enough $\delta$, is
again determined by the power-law asymptotics of  the L\'evy stable
density in its far tail: $\sigma^2 \simeq \text{const} \times
\varepsilon^\alpha \delta^{2-\alpha}\Delta t $. The variance
introduced by the background noise over the unit time interval (the
background noise strength) due to the role of summation of variances
for the white noise behaves as
\begin{equation}
\sigma^2 \simeq \text{const} \times \varepsilon^\alpha \delta^{2-\alpha}.
\label{sigma1}
\end{equation}

For $\varepsilon \to 0$ it is always possible to chose $\delta$ in such
a way that both $P$ in Eq.~(\ref{probability}) and $\sigma^2$ in
Eq.~(\ref{sigma1}) tend to zero.
In Refs.~\cite{impavl_a,impavl_b} the value of $\delta=\varepsilon^{1/2}$ was
taken effectively to obtain the necessary estimates.

Let us now fix small $\varepsilon$ and take $\delta$ as discussed
above. Between the two subsequent large spikes the particle is
subjected only to a background noise, which, for small $\varepsilon$,
is so weak that the motion of a particle is almost deterministic,
namely the particle slides towards the bottom of the potential well.
The overall structure of such motion is well seen in the bottom
panel of Fig.~\ref{fig:decomposition}. For $\varepsilon$ small enough
the background noise can be neglected, which would correspond to
smoothing the rugged curve in the bottom panel of the figure and
approximating it by a deterministic trajectory. On the other hand,
the the time lag between the two subsequent spikes is so large that
the particle reaches the bottom of the potential well between the
two spikes. The only process of escape possible in this case,
corresponds to the possibility that a strong enough spike ``kicks''
the particle from the bottom of the potential well, and an escape
occurs in a single jump if its length exceeds the distance between
the potential's minimum and maximum, see
Fig.~\ref{fig:decomposition}. The probability of jumping out of the
well $\Omega_0$ is the conditional probability of having a positive spike
larger than $(1-q)\mathcal L$ or having a negative spike
larger than $q\mathcal L$ in absolute value, that is
\begin{equation}
P_{\text{exit}}\simeq\text{Prob}\Big\{\xi(n)< -q\mathcal L\text{ or } \xi(n)> (1-q)\mathcal L \Big|\, |\xi(n)|>\delta\Big\}
= \frac{\int_{-\infty}^{-q\mathcal L} p(\xi)\, d\xi +
\int_{(1-q)\mathcal L}^{\infty} p(\xi)\, d\xi}{\int_{-\infty}^{-\delta} p(\xi)\, d\xi
+ \int_{\delta}^{\infty} p(\xi)\, d\xi}
, \label{escapeprob}
\end{equation}
where the wings of the L\'evy stable PDF representing the action of
spikes per unit time are given by
$p(\xi) \simeq A_\pm w^{\alpha}\xi^{-1-\alpha}$, $\xi\to\pm\infty$. Thus, we get
\begin{equation}
P_{\text{exit}} \simeq
\Big(\frac{A_-}{q^\alpha}+\frac{A_+}{(1-q)^\alpha}\Big)\frac{\delta^{\alpha}}{A \mathcal{L}^\alpha},
\end{equation}
which reproduces the result of Ref.~\cite{ESAIM}.

Now we are able to calculate the mean exit time from a single well. Indeed,  exit on a $k$-th spike means that the first $(k-1)$ attempts
were unsuccessful, ant the $k$-th jump was big enough.
The probability of this event approximately equals $(1-P_{\text{exit}})^{k-1}P_{\text{exit}}$. The mean
exit time in this case equals $kT$ where the mean interspike time $T$ is known, see
Eq.~\eqref{mean time}. This yields
\begin{equation}
\begin{aligned}
\label{eq:meanexittime}
\langle\tau(\varepsilon) \rangle &\simeq \sum_{k=1}^\infty kT (1-P_{\text{exit}})^{k-1}P_{\text{exit}}=\frac{T}{P_{\text{exit}}}\\
&=\left[  \frac{c}{2\alpha
\cos\frac{\pi\alpha}{2} |\Gamma(-\alpha) |} \left(
\frac{1+\beta}{(1-q)^{-\alpha}} + \frac{1-\beta}{q^{-\alpha}}
\right)\right]^{-1} \times \frac{\mathcal L^\alpha}{\varepsilon^\alpha}.
\end{aligned}
\end{equation}

The picture
of escape at small $\varepsilon$ differs drastically from the Kramers picture, namely
instead of climbing up in the potential well the particle is thrown out from the
well by a single large kick.
Due to the fact that potential is translationally invariant the mean exit times $\langle \tau^i(\varepsilon)\rangle$
for all the potential wells are the same and equal to $\langle \tau(\varepsilon)\rangle$.

Now we can derive the asymptotic value of the transition probability $p_{i,i+k}$ for a particle to make a jump
from the $i$-th potential well to the ($i+k$)-th well. Assume for brevity that the particle starts at the well $\Omega_0$, $i=0$, and jumps to the
well $\Omega_k$, $k\geqslant 1$. The probability $p_k=p_{0,k}$ is then obtained similarly to Eq.~\eqref{eq:meanexittime}.
Indeed, the transition to the well $\Omega_k$ occurs if the kick $\xi(n)$ overcomes the distance $(1-q)\mathcal L+(k-1)\mathcal L$ but
is smaller than $(1-q)\mathcal L+k\mathcal L$. The probability of this event equals to
\begin{equation}
\begin{aligned}
P_{0\to k}&\simeq\text{Prob}\Big\{\xi(n)\in [(k-q)\mathcal L, (k+1-q)\mathcal L] \Big|\, |\xi(n)|>\delta\Big\}
= \frac{\int_{(k-q)\mathcal L}^{(k+1-q)\mathcal L} p(\xi)\, d\xi }{\int_{-\infty}^{-\delta} p(\xi)\, d\xi
+ \int_{\delta}^{\infty} p(\xi)\, d\xi}\\
&=\Big(\frac{1}{(k-q)^\alpha}-\frac{1}{(k+1-q)^\alpha}\Big)\frac{A_+}{A}\frac{\delta^\alpha}{\mathcal{L}^\alpha}.
\end{aligned}\label{eq:0tok}
\end{equation}
Then with the help of the formula of the total probability we get
\begin{equation}
\begin{aligned}
p_{i,i+k}=p_k&\simeq \sum_{k=1}^\infty (1-P_{\text{exit}})^{k-1}P_{0\to k}=\frac{P_{0\to k}}{P_{\text{exit}}}=
(1+\beta)
\frac{(k-q)^{-\alpha}-(k+1-q)^{-\alpha}}{(1-\beta)q^{-\alpha}+(1+\beta)(1-q)^{-\alpha}}.
\label{eq:transitionright}
\end{aligned}
\end{equation}
Analogously, considering negative jumps we calculate transition probabilities $p_k=p_{i,i+k}$ for $k\leqslant -1$ as
\begin{equation}
p_{i,i+k}=
p_{k}\simeq(1-\beta)\frac{(-k-1+q)^{-\alpha}-(-k+q)^{-\alpha}}{(1-\beta)q^{-\alpha}+(1+\beta)(1-q)^{-\alpha}}.
\label{eq:transitionleft}
\end{equation}

\subsection{Characteristics of transport \label{sec:micro}}

A particle inserted into the $i$-th potential well spends a
random time within the potential well, until it is kicked out of
the well by a large spike, as it is described in subsection~\ref{sec:decomposition}.
It can be shown that in the small noise limit $\varepsilon\to 0$, the first exit time $\tau^i(\varepsilon)$ from $i$-th potential
$\Omega_i$ is an exponentially distributed random variable with the
mean value $\langle\tau^i(\varepsilon)\rangle=\langle  \tau(\varepsilon) \rangle$, see Eq.\ \eqref{eq:meanexittime}.
The transition probability $p_{i,i+k}$ for a particle to make a jump
from the $i$-th potential well to the ($i+k$)-th
potential well is given by Eqs.~\eqref{eq:transitionright} and \eqref{eq:transitionleft}.

\subsubsection{Splitting probabilities}

In order to elucidate the
asymmetry of the first escape we use the splitting probability $\pi$
which is a probability of a first escape from a potential well to
the right
\begin{equation}
 \pi_R(\alpha,\beta,q)=\sum\limits_{k\geqslant 1}p_k=\frac{(1+\beta)(1-q)^{-\alpha}}{(1-\beta)q^{-\alpha}+(1+\beta)(1-q)^{-\alpha}},
\label{eq:spilittingright}
\end{equation}
or to the left
\begin{equation}
 \pi_L(\alpha,\beta,q)=\sum\limits_{k\leqslant -1}p_k=\frac{(1-\beta)q^{-\alpha}}{(1-\beta)q^{-\alpha}+(1+\beta)(1-q)^{-\alpha}}.
\label{eq:spilittingleft}
\end{equation}
Due to the asymmetry of the noise it is possible to find such a set
of parameters for which the two values of splitting
probabilities are equal. The balance condition
\begin{equation}
 \pi_L(\alpha,\beta,q)=\pi_R(\alpha,\beta,q)=\frac{1}{2}
\end{equation}
leads to  the skewness parameter
\begin{equation}
 \beta_\pi(\alpha,q)=\frac{(1-q)^{\alpha}- q^{\alpha}}{(1-q)^{\alpha}+ q^{\alpha}}.
\label{eq:betapi}
\end{equation}

\begin{figure}[ht]
\begin{center}
 \includegraphics[width=0.9\columnwidth]{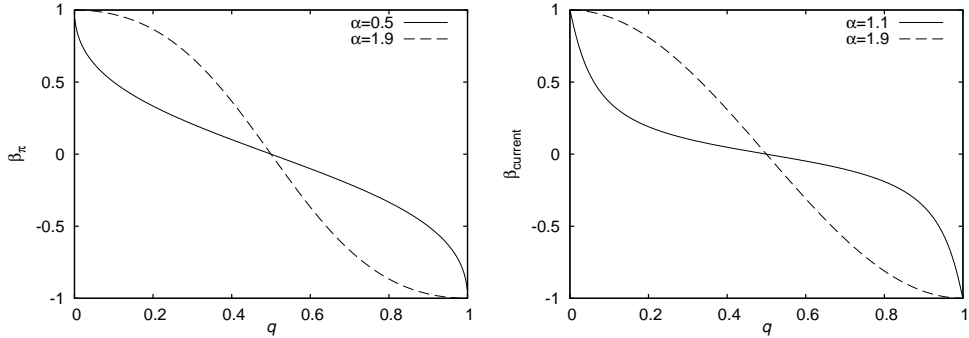}
\caption{The left panel presents values of the noise asymmetry $\beta_\pi$ leading to equal values of splitting probabilities,
see Eq.~(\ref{eq:betapi}).
The right panel presents values of the noise asymmetry $\beta_{\text{current}}$ leading to zero current, see Eq.~\eqref{eq:betacur}.
} \label{fig:beta_pi_cur}
\end{center}
\end{figure}

\subsubsection{Mean displacement}

For $\alpha>1$, the mean displacement of the particle at the time $t$ is
\begin{equation}
 \langle X^\varepsilon(t)\rangle= \langle n \rangle \Lambda,
 \label{eq:meandisplacement}
\end{equation}
where $\langle n \rangle = t/\langle \tau(\varepsilon) \rangle$ is the average number of
jumps during time $t$.
The mean value of the displacement $\Lambda$ can be expressed as
\begin{equation}
 \Lambda =\mathcal{L} \sum_{k=-\infty,k\neq 0}^\infty k p_k= W_+(\alpha,\beta,q)-W_-(\alpha,\beta,q),
\label{eq:wpwm}
\end{equation}
where
\begin{eqnarray}
\label{eq:wright}
 W_+(\alpha,\beta,q) & = &\mathcal{L}  \sum\limits_{k\geqslant 1}kp_k =\mathcal{L}
\frac{(1+\beta)\zeta(\alpha,1-q)}{(1-\beta)q^{-\alpha}+(1+\beta)(1-q)^{-\alpha}}
\nonumber
\end{eqnarray}
and
\begin{eqnarray}
\label{eq:wleft}
 W_-(\alpha,\beta,q) & = & -\mathcal{L} \sum\limits_{k\leqslant -1}kp_k
=\mathcal{L}
\frac{(1-\beta)\zeta(\alpha,q)}{(1-\beta)q^{-\alpha}+(1+\beta)(1-q)^{-\alpha}}.
\nonumber
\end{eqnarray}
In above equations $\zeta(\alpha,q)$ denotes the Hurwitz zeta
function, $\zeta(\alpha,q)=\sum_{k=0}^\infty (k+q)^{-\alpha}$.
The mean jump length $\Lambda$, see Eq.~(\ref{eq:meandisplacement}), and consequently the mean displacement $\langle X^\varepsilon (t) \rangle$, see Eq.~(\ref{eq:wpwm}), are finite for $\alpha>1$ only.

The main characteristics of the particles' transport in the L\'evy ratchet is
the particle current. For the stability index $\alpha>1$, the
current is defined as the time derivative of $\langle
X^\varepsilon(t)\rangle$,
\begin{eqnarray}
 j(\alpha,\beta,q) &=& \frac{d}{dt}\langle X^\varepsilon(t)\rangle
=\frac{W_+(\alpha,\beta,q)-W_-(\alpha,\beta,q)}{\langle \tau(\varepsilon) \rangle}. \nonumber
\label{eq:current}
\end{eqnarray}
Consequently, for small noise intensities $\varepsilon\to 0$ we get
\begin{equation}
 j(\alpha,\beta,q) \propto \varepsilon^\alpha.
\label{eq:currentscalling}
\end{equation}
The multiplicative constant in Eq.~(\ref{eq:currentscalling}) depends on parameters  $\alpha$ and $\beta$ of the noise and on
the asymmetry parameter $q$ of the potential.

We also determine the skewness parameter $\beta$ of the noise, which leads to the zero particles' current. Indeed, the balance equation
\begin{equation}
 j(\alpha,\beta,q)=0
\end{equation}
yields the unique solution
\begin{equation}
\label{eq:betacur}
\beta_{\text{current}}(\alpha,q)=\frac{\zeta(\alpha,q)-\zeta(\alpha,1-q) }{\zeta(\alpha,q)+\zeta(\alpha,1-q) }.
\end{equation}

\section{Numerical validation of the theory\label{sec:results}}

\begin{figure}[ht]
\begin{center}\resizebox{0.9\columnwidth}{!}{
\includegraphics{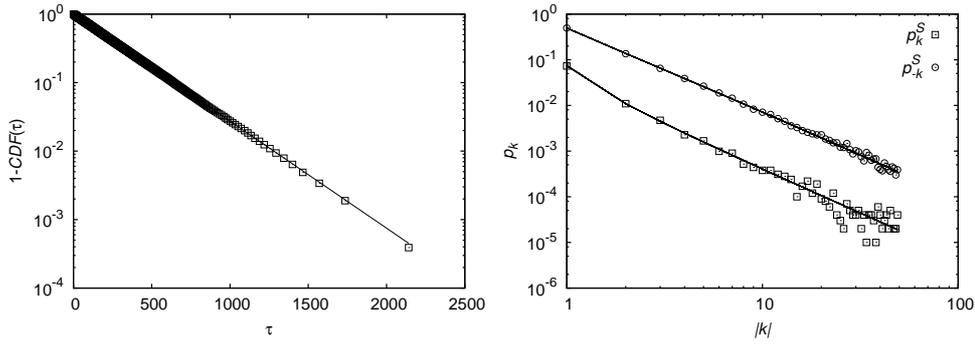}}
\caption{
The left panel demonstrates the exponential character of the (first)
exit time distribution.
The solid line in the left panel presents $\exp[-\tau/\langle \tau \rangle^T]$. The right panel presents transition probabilities $p_{k}$ from
the initial potential well $(0)$ to the final potential well ($k$).
Circles and squares represent simulation results, while thin solid lines
correspond to analytical formulas given by
Eqs.~(\ref{eq:transitionright}) and (\ref{eq:transitionleft}).
 Simulation parameters: the stability index $\alpha=0.9$, the noise asymmetry
$\beta=-0.9$, the potential asymmetry $q=0.7$ and the noise intensity $\varepsilon=2^{-9}$.}
\label{fig:distributions}
\end{center}
\end{figure}

\begin{figure}[ht]
\begin{center}\resizebox{0.5\columnwidth}{!}{
\includegraphics{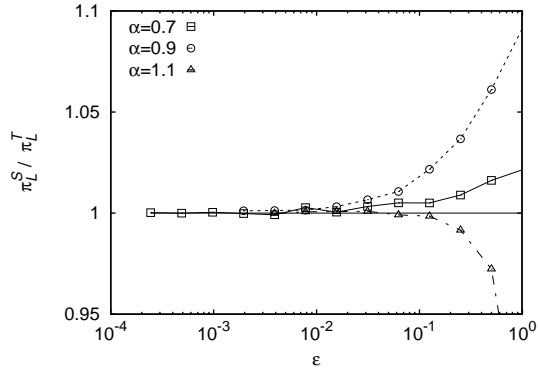}}
\caption{Ratio of simulated and theoretical values of splitting probabilities $\pi^S_L/\pi^T_L$.
Simulation parameters: the noise asymmetry $\beta=-0.9$, the potential asymmetry $q=0.7$.}
\label{fig:averages_pi}
\end{center}
\end{figure}

\begin{figure}[ht]
\begin{center}\resizebox{0.9\columnwidth}{!}{
\includegraphics{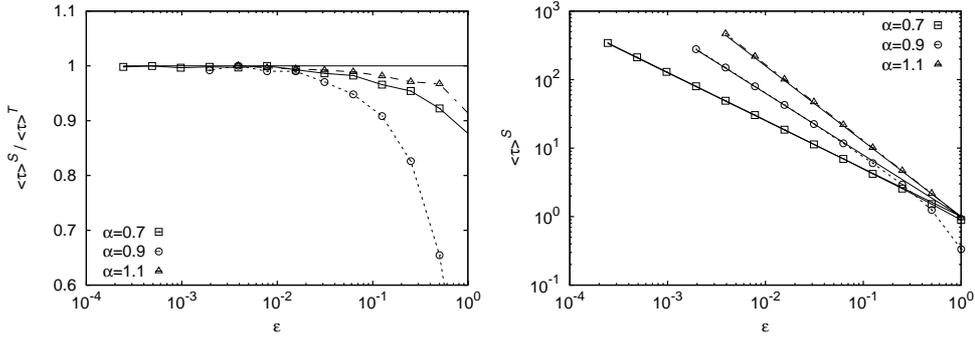}}
\caption{
The left panel demonstrates the ratio of simulated and theoretical values of the mean escape time
$\langle \tau \rangle^S/\langle \tau \rangle^T$.  The right panel presents the power-law scaling of the mean escape time as a function of the noise intensity
$\varepsilon$. Points represent simulation results while
solid lines indicate the theoretical scaling of
the mean exit time $\langle \tau \rangle$ with the noise intensity $\varepsilon$. Simulation parameters:
the noise asymmetry $\beta=-0.9$, the potential asymmetry $q=0.7$.}
\label{fig:averages_p}
\end{center}
\end{figure}

\begin{figure}[ht]
\begin{center}\resizebox{0.5\columnwidth}{!}{
\includegraphics{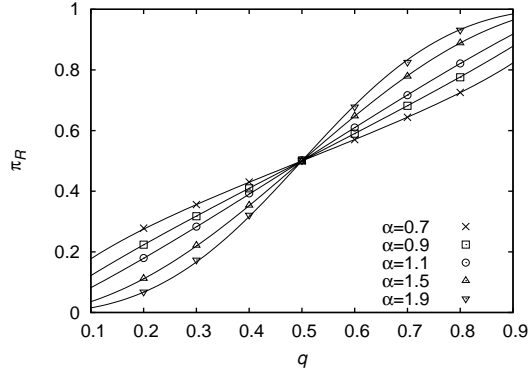}}
\caption{Theoretical (solid lines) and simulated (points) values of the splitting probability $\pi_R$ as a function
of the potential asymmetry $q$. Various curves correspond to different values of the stability index $\alpha$. Simulation parameters:
the noise asymmetry $\beta=0.0$, the noise intensity $\varepsilon=2^{-9}$ and the potential asymmetry $q=0.7$.}
\label{fig:splitww_b00}
\end{center}
\end{figure}

\begin{figure}[ht]
\begin{center}\resizebox{0.95\columnwidth}{!}{
\includegraphics{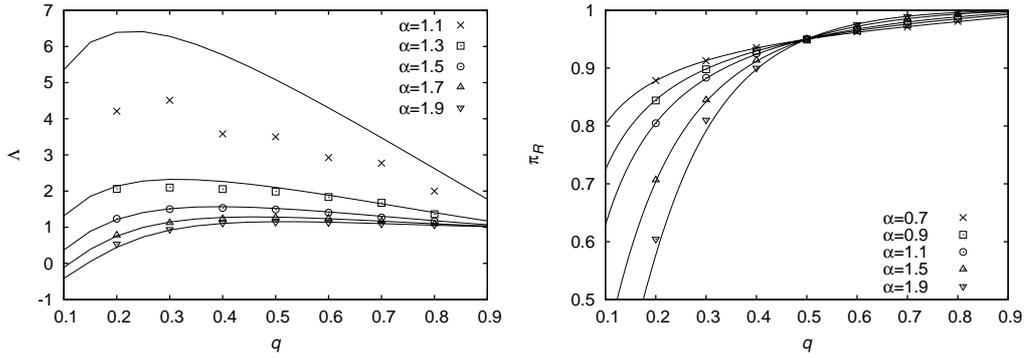}}
\caption{
Theoretical (solid lines) and simulated (points) values of the mean displacement $\Lambda^S=\langle \Lambda_i \rangle$
(left panel) and theoretical (solid lines) and simulated  (points) values of the splitting probability $\pi_R$ (right panel).
Simulation parameters:
the noise asymmetry $\beta=0.9$, the noise intensity $\varepsilon=2^{-9}$ and the potential asymmetry $q=0.7$.}
\label{fig:splitww}
\end{center}
\end{figure}

\begin{figure}[ht]
\begin{center}
\includegraphics[width=0.95\columnwidth]{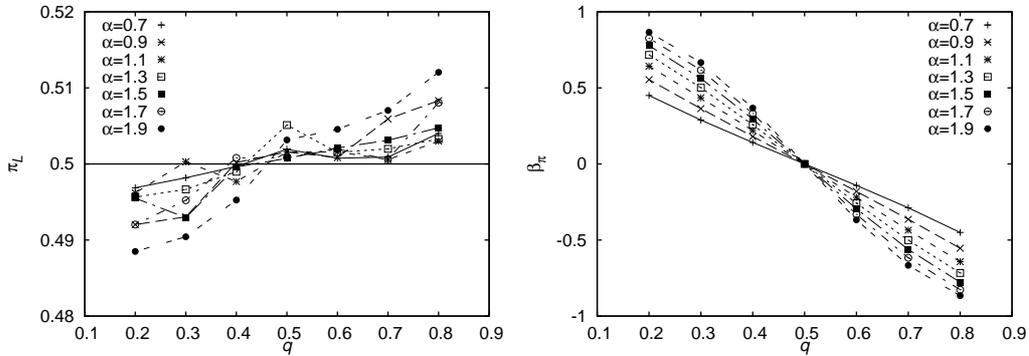}
\caption{The left panel presents the simulated  spitting probability $\pi_L$ for the set of parameters for which the
balance equality $\pi_L=\pi_R=0.5$ should
hold. The right panel presents values of the noise asymmetry $\beta_\pi$ leading to equal values of splitting
probabilities, see Eq.~(\ref{eq:betapi}). Simulation parameters: $\varepsilon=2^{-6}$.}
\label{fig:pistop}
\end{center}
\end{figure}

\begin{figure}[ht]
\begin{center}
\includegraphics[width=0.95\columnwidth]{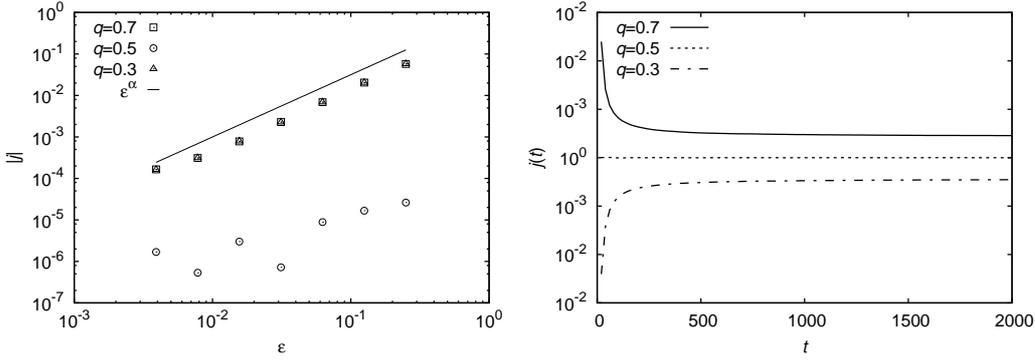}
\caption{The left
panel presents the current $j$ as a function of the noise intensity
$\varepsilon$ at $t=2000$ and various values of the potential
asymmetry. The solid line presents theoretical scaling of the
current on $\varepsilon$.
The right panel represents the current $j$ as a function of time for the
symmetric L\'evy noise and various values of potential asymmetry
$q$. Simulation parameters: $\alpha=1.5$, $\beta=0$, and $\varepsilon=2^{-5}$.
} \label{fig:a17_b00_current}
\end{center}
\end{figure}

\begin{figure}[ht]
\begin{center}
\includegraphics[width=0.95\columnwidth]{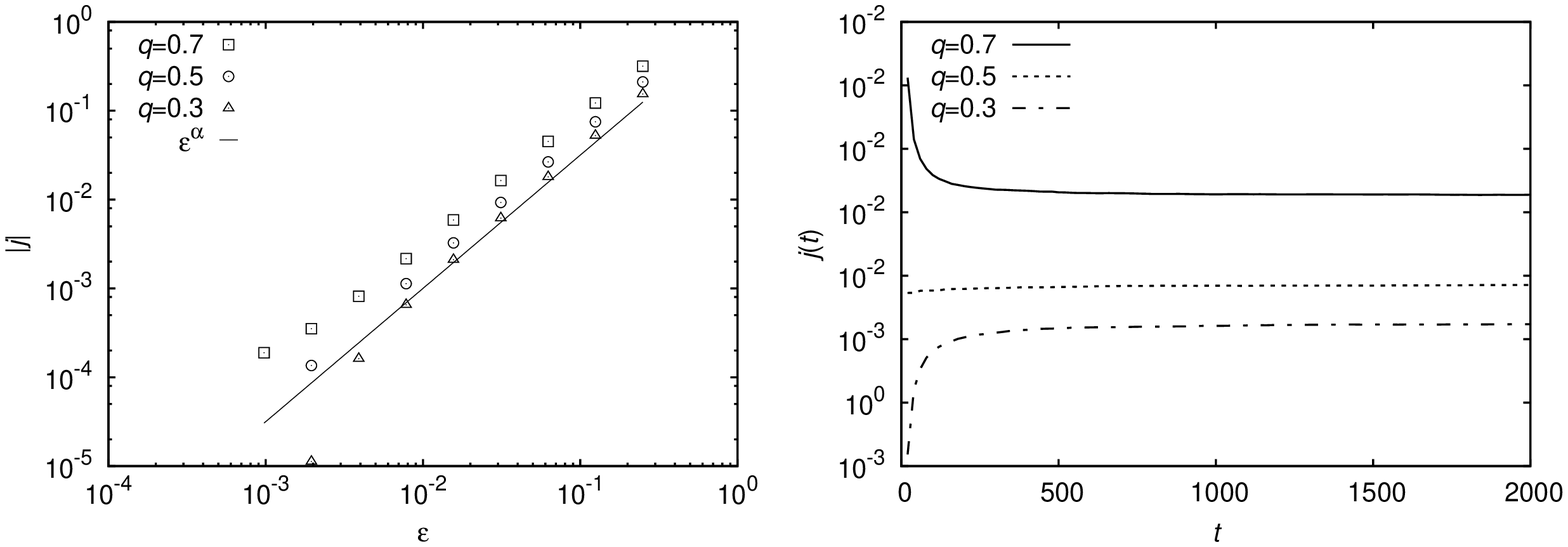}
\caption{The same as in Fig.~\ref{fig:a17_b00_current} for
asymmetric L\'evy noise with $\beta=0.9$. Simulation parameters: $\alpha=1.5$,
$\varepsilon=2^{-6}$.} \label{fig:a15_b09_current}
\end{center}
\end{figure}

In order to show validity of the developed theory extensive
numerical simulations have been performed for the stability index
$\alpha\in\{0.7,0.9,1.1,1.5,1.9\}$, the noise asymmetry parameter
$\beta\in\{0.0, \pm 0.5, \pm 0.9\}$ and the potential asymmetry
parameter $q\in\{0.2, 0.3, 0.4, 0.5, 0.6, 0.7, 0.8\}$ with various
(decreasing) noise intensities $\varepsilon$. Numerical results were
constructed by standard methods of integration of stochastic
differential equations with respect to $\alpha$-stable noises, see
Refs. \cite{janicki1994,janicki1996,dybiec2004b,dybiec2006}. All
numerical simulations have been performed with the time step of
integration $\Delta t=10^{-3}$, the number of realizations $N=10^5$,
the potential depth $V_0=1$, the scale parameter $c=1$ and the
potential segment length $\mathcal{L}=1$. In general, all
simulations in the limit of weak noise (small $\varepsilon$)
converge to theoretical predictions. From the whole set of
simulations we have chosen exemplary results presented in
Figs.~\ref{fig:distributions}--\ref{fig:a15_b09_current}.

Figure~\ref{fig:distributions} presents the complementary cumulative distribution of
the exit time $1-CDF(\tau)$ (left panel) and the transition probabilities
$p_{k}$ (right panel) for the stability index
$\alpha=0.9$, the noise asymmetry $\beta=-0.9$ and the potential
asymmetry $q=0.7$. The left panel of Fig.~\ref{fig:distributions}
demonstrates the exponential character of the exit time
distribution whereas the right panel of Fig.~\ref{fig:distributions}
compares theoretical transition probabilities from the initial potential well $\Omega_0$
to the well $\Omega_k$ given by
Eqs.~(\ref{eq:transitionright}) and (\ref{eq:transitionleft}) (solid
lines) with their numerical estimators (symbols), demonstrating
perfect agreement between theoretical predictions and numerical
simulations.

Fig.~\ref{fig:averages_pi} demonstrates the ratio of
the  numerically estimated values of the splitting probabilities
$\pi^S_L$ and their theoretical values $\pi_L^T$ given by
Eq.~(\ref{eq:spilittingleft}) for different values of  $\alpha$. In
the limit of the small noise intensity the ratio between estimated
and theoretical values of splitting probabilities tends to 1
indicating the agreement between the theory and simulations. For
larger noise intensities numerical results deviate from predictions
for weak noise regime. The sign of this deviation is different for
$\alpha>1$ and for $\alpha<1$.

The left panel of
Fig.~\ref{fig:averages_p} displays the ratio between the numerically
estimated mean exit time $\langle \tau(\varepsilon) \rangle^S$ and
the theoretical prediction $\langle \tau(\varepsilon) \rangle ^T$,
see Eq.~(\ref{eq:meanexittime}). In the limit of small noise
intensities $\varepsilon$ the ratio between estimated and
theoretical values of mean exit times tends to 1 indicating that
simulations performed corroborate theoretical findings.
The right panel of Fig.~\ref{fig:averages_p} shows the scaling of the
estimated mean exit time $\langle \tau(\varepsilon) \rangle$
(symbols) with the noise intensity $\varepsilon$ along with the
corresponding theoretical prediction, see
Eq.~(\ref{eq:meanexittime}), shown as solid lines. Deviations from
the predicted scaling visible for large noise intensities vanish for
small values of $\varepsilon$.

Fig.~\ref{fig:splitww_b00} displays the splitting probability
$\pi_R$ given by Eq.~(\ref{eq:spilittingright}) as a function of
potential asymmetry $q$ for the case of symmetric L\'evy noise with
$\beta=0$ and different $\alpha$. Here again the close agreement
between analytical formulas and numerical results is observed. For
symmetric noise and symmetric potential ($q=0.5$) splitting
probability is equal to $0.5$. This indicates absence of the current
in the system at hand.

The left panel of Fig.~\ref{fig:splitww} presents the mean
displacement $\Lambda$ defined for $\alpha>1$ only and given by Eq.~(\ref{eq:wpwm}),
and its numerical value estimated as
\begin{equation}
\Lambda^{S} = \langle \Lambda_i \rangle,
\end{equation}
$\Lambda_i$ being the number of the potential well to which a
random walker was thrown from the initial one. For all values of
$\alpha$ except for $\alpha = 1.1$, which is too close to the
boundary of convergence, the agreement between the numerical and the
theoretical values is good.
The right panel of Fig.~\ref{fig:splitww} presents the splitting
probability $\pi_R$ as a function of the potential asymmetry $q$
for the case of asymmetric L\'evy noise with $\beta=0.9$. The
numerical values (symbols) agree  well with theoretical predictions
(lines).

Fig.~\ref{fig:pistop} demonstrates the value of the splitting probability $\pi_L$ observed for the value of
 $\beta_\pi$ given by Eq.~(\ref{eq:betapi}) (left panel)
and values
of the noise asymmetry parameter $\beta_\pi$ leading to equal values of splitting
probabilities, see Eq.~(\ref{eq:betapi}), (right panel). Results presented in Fig.~\ref{fig:pistop} correspond to $\varepsilon=2^{-6}$,
which is larger than $\varepsilon$ in remaining figures. This is due to the fact that for noise asymmetry $\beta_\pi$ given by
Eq.~(\ref{eq:betapi}) longer simulation time is necessary.

The left panels of Figs.~\ref{fig:a17_b00_current} and
\ref{fig:a15_b09_current} compare theoretical values of the current
given by Eq.~(\ref{eq:current}) with their numerical estimates as
functions  of the noise intensity $\varepsilon$ at time $t=2000$.
The power-law dependence $j\propto
\varepsilon^\alpha$ of the current, see Eq.~(\ref{eq:currentscalling}) is clearly demonstrated.
For symmetric noises, see Fig.~\ref{fig:a17_b00_current}, the
absolute values of the current for $q=0.3$ and $q=0.7$ are the same.
For $q=0.5$, the theoretical value of the current vanishes,
while numerical results fluctuate at a very low level.
The right panels of Figs.~\ref{fig:a17_b00_current} and
\ref{fig:a15_b09_current} present the dependence of the current $j$
on time $t$ for symmetric ($\beta=0$) and asymmetric ($\beta=0.9$)
noise, respectively. Various curves correspond to different values
of the potential asymmetry $q$. After a short transient the current
reaches a constant value corresponding to a stationary regime of
operation.

\section{Summary\label{sec:summary}}

In this paper we have considered the transport properties of a particle
embedded in a ratchet potential and subjected to a white L\'evy
noise. In the case of weak noise the decomposition method proposed
in Refs.~\cite{impavl_a,impavl_b} allowed us to calculate such
characteristics of this transport as splitting probabilities of the
first escape from a single well of a ratchet potential, the
transition probability, the mean displacement of a particle and the
particle current. In order to confirm analytical results, we
performed an extensive numerical simulations demonstrating good
agreement with the theory at small values of the noise intensity.


\begin{acknowledgement}
The authors acknowledge support by DFG within SFB555. AVC
acknowledges financial support from European Commission via MC IIF,
grant 219966 LeFrac. Computer simulations have been performed at
Institute of Physics, Jagellonian University and Academic Computer
Center, Cyfronet AGH.
\end{acknowledgement}



\begin{thebibliography}{10}

\bibitem{levy}
P.~P. L{\'e}vy.
\newblock {\em Th{\'e}orie de {l'a}ddition des variables al{\'e}atoires}.
\newblock Gauthier-Villars, Paris, 1937.

\bibitem{mandelbrot}
B.~Mandelbrot.
\newblock {\em The Fractal Geometry of Nature}.
\newblock W. H. Freeman, New York, 1983.

\bibitem{feller}
W.~Feller.
\newblock {\em {An introduction to probability theory and its applications:
  volume II}}.
\newblock John Wiley \& Sons, 1971.

\bibitem{gnedenko-kolmogorov}
B.~V. Gnedenko and A.~N. Kolmogorov.
\newblock {\em Limit distributions for sums of independent random variables}.
\newblock Addison-Wesley Publishing Company, Cambridge, Mass., 1954.

\bibitem{samorod}
G.~Samorodnitsky and M.~S. Taqqu.
\newblock {\em Stable non-{G}aussian random processes}.
\newblock {Chapman\&Hall/CRC}, 1994.

\bibitem{nature93}
M.~F. Shlesinger, G.~M. Zaslavsky, and J.~Klafter.
\newblock Strange kinetics.
\newblock {\em Nature}, 363(6424):31--37, 1993.

\bibitem{phystoday96}
J.~Klafter, M.~F. Shlesinger, and G.~Zumofen.
\newblock Beyond {B}rownian motion.
\newblock {\em Physics Today}, 49(2):33, 1996.

\bibitem{swinney}
T.~H. Solomon, E.~R. Weeks, and H.~L. Swinney.
\newblock Observation of anomalous diffusion and {L{\'e}vy} flights in a
  two-dimensional rotating flow.
\newblock {\em Physical Review Letters}, 71(24):3975----3978, 1993.

\bibitem{diego1}
D.~del Castillo-Negrete.
\newblock Asymmetric transport and non-{G}aussian statistics of passive scalars
  in vortices in shear.
\newblock {\em Physics of Fluids}, 10(3):576--594, 1998.

\bibitem{chechkinplasma}
A.~V. Chechkin, V.~Yu. Gonchar, and M.~Szyd{\l}owski.
\newblock Fractional kinetics for relaxation and superdiffusion in a magnetic
  field.
\newblock {\em Physics of Plasmas}, 9(1):78--88, 2002.

\bibitem{goncharplasma}
V.~Yu. Gonchar, A.~V. Chechkin, E.~L. Sorokovoi, V.~V. Chechkin, and E.~D.
  Volkov.
\newblock Stable {L{\'e}vy} distributions of the density and potential
  fluctuations in the edge plasma of the {$U$-$3M$} torsatron.
\newblock {\em Plasma Physics Reports}, 29(5):380--390, 2003.

\bibitem{goncharplasma1}
T.~Mizuuchi, V.~V. Chechkin, K.~Ohashi, E.~L. Sorokovoy, A.~V. Chechkin, V.~Yu.
  Gonchar, K.~Takahashi, S.~Kobayashi, H.~Okada {K. Nagasaki}, S.~Yamamoto,
  F.~Sano, K.~Kondo, N.~Nishino, H.~Kawazome, H.~Shidara, S.~Kaneko,
  Y.~Fukagawa, Y.~Morita, S.~Nakazawa, S.~Nishio, S.~Tsuboi, and M.~Yamada.
\newblock Edge fluctuation studies in {H}eliotron {J}.
\newblock {\em Journal of Nuclear Materials,}, 337--339:332--336, 2005.

\bibitem{diego2}
D.~del Castillo-Negrete, B.~A. Carreras, and V.~E. Lynch.
\newblock Nondiffusive transport in plasma turbulence: A fractional diffusion
  approach.
\newblock {\em Physical Review Letters}, 94(6):065003, 2005.

\bibitem{walther}
H.~Katori, S.~Schlipf, and H.~Walther.
\newblock Anomalous dynamics of a single ion in an optical lattice.
\newblock {\em Physical Review Letters}, 79(12):2221--2224, 1997.

\bibitem{galgani}
A.~Carati, L.~Galgani, and B.~Pozzi.
\newblock {L{\'e}vy} flights in the {L}andau-{T}eller model of molecular
  collisions.
\newblock {\em Physical Review Letters}, 90(1):010601, 2003.

\bibitem{peng}
C.-K. Peng, J.~Mietus, J.~M. Hausdorff, S.~Havlin, H.~E. Stanley, and A.~L.
  Goldberger.
\newblock Long-range anticorrelations and non-{G}aussian behavior of the
  heartbeat.
\newblock {\em Physical Review Letters}, 70(9):1343--1346, 1993.

\bibitem{segev}
R.~Segev, M.~Benveniste, E.~Hulata, N.~Cohen, A.~Palevski, E.~Kapon,
  Y.~Shapira, and E.~Ben-Jacob.
\newblock Long term behavior of lithographically prepared \textit{in vitro}
  neuronal networks.
\newblock {\em Physical Review Letters}, 88(11):118102, 2002.

\bibitem{lomholt}
M.~A. Lomholt, T.~Ambj{\"o}rnsson, and R.~Metzler.
\newblock Optimal target search on a fast folding polymer chain with volume
  exchange.
\newblock {\em Physical Review Letters}, 95(26):260603, 2005.

\bibitem{buldyrev}
G.~M. Viswanathan, V.~Afanasyev, S.~V. Buldyrev, E.~J. Murphey, P.~A. Prince,
  and H.~E. Stanley.
\newblock L{\'e}vy flight search patterns of wandering albatrosses.
\newblock {\em Nature}, 381:413--415, 1996.

\bibitem{vidal}
O.~Sotolongo-Costa, J.~C. Antoranz, A.~Posadas, F.~Vidal, and A.~V{\'a}zquez.
\newblock L{\'e}vy flights and earthquakes.
\newblock {\em Geophysical Research Letters}, 27(13):1965--1968, 2000.

\bibitem{ditlevsen}
P.~D. Ditlevsen.
\newblock {Observation of $\alpha$-stable noise induced millenial climate
  changes from an ice record}.
\newblock {\em Geophysical Research Letters}, 26(10):1441--1444, May 1999.

\bibitem{kassam}
S.~A. Kassam.
\newblock {\em Signal Detection in Non-Gaussian Noise}.
\newblock Springer Texts in Electrical Engineering. Springer, 1988.

\bibitem{WegmanTS88}
E.~J. Wegman, J.~B. Thomas, and S.~C. Schwartz, editors.
\newblock {\em Topics in Non-Gaussian Signal Processing}.
\newblock Springer-Verlag, 1988.

\bibitem{NikiasS95}
C.~L. Nikias and M.~Shao.
\newblock {\em Signal processing with alpha-stable distributions and
  applications}.
\newblock Wiley-Interscience, New York, NY, USA, 1995.

\bibitem{MantegnaS99}
R.~N. Mantegna and H.~E. Stanley.
\newblock {\em An Introduction to Econophysics: Correlations and Complexity in
  Finance}.
\newblock Cambridge University Press, 1999.

\bibitem{BouchaudP00}
J.-P. Bouchaud and M.~Potters.
\newblock {\em Theory of Financial Risks: From Statistical Physics to Risk
  Management}.
\newblock Cambridge University Press, 2000.

\bibitem{GomoryM97}
R.~E. Gomory and B.~B. Mandelbrot.
\newblock {\em Fractals and Scaling In Finance: Discontinuity, Concentration,
  Risk}.
\newblock Springer, 1997.

\bibitem{brockmann}
D.~Brockmann, L.~Hufnagel, and T.~Geisel.
\newblock The scaling laws of human travel.
\newblock {\em Nature}, 439:462--465, 2006.

\bibitem{barthelemy}
P.~Barthelemy, J.~Bertolotti, and D.~S. Wiersma.
\newblock A {L{\'e}vy} flight for light.
\newblock {\em Nature}, 453:495--498, 2008.

\bibitem{mercadier}
N.~Mercadier, W.~Guerin, M.~Chevrollier, and R.~Kaiser.
\newblock L{\'e}vy flights of photons in hot atomic vapours.
\newblock {\em Nature Physics}, 5:602--605, 2009.

\bibitem{Jespersen}
S.~Jespersen, R.~Metzler, and H.~C. Fogedby.
\newblock {L{\'e}vy flights in external force fields: Langevin and fractional
  Fokker-Planck equations and their solutions}.
\newblock {\em Physical Review E}, 59(3):2736--2745, 1999.

\bibitem{CheGo}
A.~V. Chechkin and V.~Yu. Gonchar.
\newblock Linear relaxation processes governed by fractional symmetric kinetic
  equations.
\newblock {\em Journal of Experimental and Theoretical Physics},
  91(3):635--651, 2000.

\bibitem{chemphys}
A.~V. Chechkin, V.~Yu. Gonchar, J.~Klafter, R.~Metzler, and L.~V. Tanatarov.
\newblock Stationary states of non-linear oscillators driven by {L{\'e}vy}
  noise.
\newblock {\em Chemical Physics}, 284(1-2):233--251, 2002.

\bibitem{SamorodnitskyG-03}
G.~Samorodnitsky and M.~Grigoriu.
\newblock Tails of solutions of certain nonlinear stochastic differential
  equations driven by heavy tailed {L}{\'e}vy motions.
\newblock {\em Stochastic Processes and their Applications}, 105(1):69--97,
  2003.

\bibitem{SamGri07}
G.~Samorodnitsky and M.~Grigoriu.
\newblock Characteristic function for the stationary state of a one-dimensional
  dynamical system with {L{\'e}vy} noise.
\newblock {\em Theoretical and Mathematical Physics}, 150(3):332--346, 2007.

\bibitem{dubkov}
A.~Dubkov and B.~Spagnolo.
\newblock {L}angevin approach to {L{\'e}vy} flights in fixed potentials: Exact
  results for stationary probability distributions.
\newblock {\em Acta Physica Polonica B}, 38(5):1745--1758, 2007.

\bibitem{chekla}
A.~V. Chechkin, J.~Klafter, V.~Yu. Gonchar, R.~Metzler, and L.~V. Tanatarov.
\newblock Bifurcation, bimodality, and finite variance in confined {L{\'e}}vy
  flights.
\newblock {\em Physical Review E}, 67:010102(R), 2003.

\bibitem{ChechkinGKM-04}
A.~V. Chechkin, V.~Yu. Gonchar, J.~Klafter, R.~Metzler, and L.~V. Tanatarov.
\newblock {L{\'e}vy} flights in a steep potential well.
\newblock {\em Journal of Statistical Physics}, 115(5--6):1505--1535, 2004.

\bibitem{Ditlevsen-04}
O.~Ditlevsen.
\newblock {Invalidity of the spectral {Fokker--Planck} equation for {Cauchy}
  noise driven {Langevin} equation}.
\newblock {\em Probabilistic Engineering Mechanics}, 19(4):385--392, 2004.

\bibitem{JStat}
B.~Dybiec, I.~M. Sokolov, and A.~V. Chechkin.
\newblock Stationary states in single-well potentials under symmetric
  {L{\'e}vy} noises.
\newblock {\em Journal of Statistical Mechanics: Theory and Experiment},
  P07008, 2010.

\bibitem{ditlevsen_2}
P.~D. {Ditlevsen}.
\newblock {Anomalous jumping in a double-well potential}.
\newblock {\em Physical Review E}, 60(1):172--179, 1999.

\bibitem{chegoslyu}
A.~V. Chechkin, V.~Yu. Gonchar, J.~Klafter, and R.~Metzler.
\newblock Barrier crossings of a {L}{\'e}vy flight.
\newblock {\em Europhysics Letters}, 72(3):348--354, 2005.

\bibitem{ChechkinSMK-07}
A.~Chechkin, O.~Sliusarenko, R.~Metzler, and J.~Klafter.
\newblock Barrier crossing driven by {L{\'e}}vy noise: Universality and the
  role of noise intensity.
\newblock {\em Physical Review E}, 75:041101, 2007.

\bibitem{bartek_1}
B.~Dybiec, E.~Gudowska-Nowak, and P.~H{\"a}nggi.
\newblock Escape driven by $\alpha$-stable white noises.
\newblock {\em Physical Review E}, 75(2):021109, 2007.

\bibitem{impavl_a}
P.~{Imkeller} and I.~{Pavlyukevich}.
\newblock {L}{\'e}vy flights: transitions and meta--stability.
\newblock {\em Journal of Physics A: Mathematical and General}, 39:L237--L246,
  2006.

\bibitem{impavl_b}
P.~{Imkeller} and I.~{Pavlyukevich}.
\newblock {First exit times of SDEs driven by stable {L}{\'e}vy processes}.
\newblock {\em Stochastic Processes and their Applications}, 116(4):611--642,
  2006.

\bibitem{ESAIM}
P.~Imkeller and I.~Pavlyukevich.
\newblock {Metastable behaviour of small noise {L{\'e}}vy--driven diffusions}.
\newblock {\em ESAIM: Probaility and Statistics}, 12:412--437, 2008.

\bibitem{hartley}
H.~Szu and R.~Hartley.
\newblock Fast simulated annealing.
\newblock {\em Physics Letters A}, 122(3,4):157--162, 1987.

\bibitem{pavl_jpa2007}
I.~Pavlyukevich.
\newblock {Cooling down L{\'e}vy flights}.
\newblock {\em Journal of Physics A: Mathematical and Theoretical},
  40:12299--12313, 2007.

\bibitem{pavl_jcp2007}
I.~Pavlyukevich.
\newblock {L{\'e}vy flights, non-local search and simulated annealing}.
\newblock {\em Journal of Computational Physics}, 226(2):1830--1844, 2007.

\bibitem{diegoche}
D.~del Castillo-Negrete, V.~Yu. Gonchar, and A.~V. Chechkin.
\newblock Fluctuation-driven directed transport in the presence of {L{\'e}vy}
  flights.
\newblock {\em Physica A}, 387(27):6693--6704, 2008.

\bibitem{dybiec2008e}
B.~Dybiec.
\newblock Current inversion in the {L{\'e}vy} ratchet.
\newblock {\em Physical Review E}, 78(6):061120, 2008.

\bibitem{dybiec2007d}
B.~Dybiec, E.~Gudowska-Nowak, and I.~M. Sokolov.
\newblock Stationary states in {L}angevin dynamics under asymmetric {L{\'e}vy}
  noises.
\newblock {\em Physical Review E}, 76:041122, 2007.

\bibitem{dybiec2007e}
B.~Dybiec and E.~Gudowska-Nowak.
\newblock Bimodality and hysteresis in systems driven by confined {L{\'e}vy}
  flights.
\newblock {\em New Journal of Physics}, 9:452, 2007.

\bibitem{reimann}
P.~Reimann.
\newblock Brownian motors: noisy transport far from equilibrium.
\newblock {\em Physics Reports}, 361(2-4):57--265, 2002.

\bibitem{bartek_2}
B.~Dybiec, E.~Gudowska-Nowak, and I.~M. Sokolov.
\newblock Transport in a {L{\'e}vy} ratchet: Group velocity and distribution
  spread.
\newblock {\em Physical Review E}, 78:011117, 2008.

\bibitem{benkadda}
M.~Vlad, F.~Spineanu, and S.~Benkadda.
\newblock Impurity pinch from a ratchet process.
\newblock {\em Physical Review Letters}, 96(8):085001, 2006.

\bibitem{phillips}
R.~Phillips and S.~R. Quake.
\newblock The biological frontier of physics.
\newblock {\em Physics Today}, 59:38--43, 2006.

\bibitem{zolotarev_uchaikin}
V.~V. Uchaikin and V.~M. Zolotarev.
\newblock {\em {Chance and stability. Stable distributions and their
  applications}}.
\newblock {Modern Probability and Statistics}. VSP, 1999.

\bibitem{janicki1994}
A.~Janicki and A.~Weron.
\newblock {\em Simulation and chaotic behaviour of $\alpha$-stable stochastic
  processes}, volume 178 of {\em Pure and Applied Mathematics}.
\newblock Marcel Dekker, Inc., 1994.

\bibitem{janicki1996}
A.~Janicki.
\newblock {\em Numerical and Statistical Approximation of Stochastic
  Differential Equations with {non-Gaussian} Measures}.
\newblock Hugo Steinhaus Centre for Stochastic Methods, Wroc{\l}aw, 1996.

\bibitem{dybiec2004b}
B.~Dybiec and E.~Gudowska-Nowak.
\newblock Resonant activation driven by strongly non-{G}aussian noises.
\newblock {\em Fluctuation and Noise Letters}, 4(2):L273--L285, 2004.

\bibitem{dybiec2006}
B.~Dybiec, E.~Gudowska-Nowak, and P.~H{\"a}nggi.
\newblock {L{\'e}vy}-{B}rownian motion on finite intervals: Mean first passage
  time analysis.
\newblock {\em Physical Review E}, 73(4):046104, 2006.

\end{thebibliography}

\end{document}